\documentstyle[eqsecnum,aps,epsf]{revtex} 
\newcommand{\postscript}[2] {\setlength{\epsfxsize}{#2\hsize}
\centerline{\epsfbox{#1}}}

\begin{document}

\twocolumn[\hsize\textwidth\columnwidth\hsize\csname@twocolumnfalse\endcsname

\title{Numerical study of the 
spherically-symmetric Gross-Pitaevskii equation in two space dimensions}

\author{Sadhan K. Adhikari}
\address{Instituto de F\'{\i}sica Te\'orica, Universidade Estadual
Paulista, 01.405-900 S\~ao Paulo, S\~ao Paulo, Brazil\\}

\date{\today}
\maketitle
\begin{abstract}

We present a numerical study of the time-dependent and time-independent
Gross-Pitaevskii (GP) equation in two space dimensions, which describes
the Bose-Einstein condensate of trapped bosons at ultralow temperature
with both attractive and repulsive interatomic interactions.  Both
time-dependent and time-independent GP equations are used to study the
stationary problems.  In addition the time-dependent approach is used to
study some evolution problems of the condensate.  Specifically, we study
the evolution problem where the trap energy is suddenly changed in a
stable preformed condensate. In this case the system oscillates with
increasing amplitude and does not remain limited between two stable
configurations.  Good convergence is obtained in all cases studied.

{\bf PACS Number(s): 02.70.Rw, 02.60.Lj, 03.75.Fi}

\end{abstract}

\vskip1.5pc]
 \newpage \section{Introduction}

Recent experiments \cite{1} of Bose-Einstein condensation (BEC) in dilute
bosonic atoms (alkali and hydrogen atoms)  employing magnetic traps at
ultra-low temperatures have intensified theoretical investigations on
various aspects of the condensate \cite{3,4,5,CHI,6,cst,7,12a,11,13}. The
properties of the condensate are usually described by the nonlinear
mean-field Gross-Pitaevskii (GP)  equation \cite{8}, which properly
incorporates the trap potential as well as the interaction among the
atoms.  The GP equation in both time-dependent and independent forms is
formally similar to the Schr\"odinger equation with a nonlinear term. The
effect of the interaction leads to the nonlinear term, which complicates
the solution procedure. There have been several numerical studies of the
GP equation in three space dimensions \cite{4,5,CHI,6,cst}.

A Bose gas in lower dimensions $-$ one and two dimensions $-$ exhibits
unusual features. For an ideal Bose gas BEC cannot occur in one and two
space dimensions at a finite temperature because of thermal fluctuations
\cite{12a,9}. The absence of BEC in one and two space dimensions has also
been established for interacting uniform systems \cite{9}. 
 However, condensation can take place under the action of a trap potential
\cite{12a,10} both for an ideal as well as interacting Bose gas. 

Although, there has been no experimental realization of BEC in two space
dimensions, this is a problem of great theoretical and experimental
interests. In a usual experiment of BEC in three space dimensions under
the action of a magnetic trap the typical thermal energy $k_BT_c$ is
assumed to be much larger than energy of oscillator quantum $\hbar
\omega$, where $k_B$ is the Boltzmann constant, $T_c$ the critical
temperature, and $\omega$ the oscillator frequency. This will allow
thermal oscillation in all three directions.  Usually, in a typical
experimental situation the oscillator frequencies in three different
directions, $x$, $y$, and $z$, are different. It is possible to obtain a
quasi-two-dimensional BEC in a real three-dimensional trap by choosing the
frequency in the third direction $\omega _z$ to satisfy $\hbar \omega_z >
k_B T_c > \hbar \omega_x, \hbar \omega_y$. In that case the energy for
thermal fluctuation is much smaller than the oscillator energy in the $z$
direction. Consequently, any motion in the $z$ direction will be frozen
and this will lead to a realization of BEC in two space dimensions. The
main features of BEC in two dimensions under the action of a harmonic trap
has been discussed by Mullin recently \cite{12a}. Also, there has been
consideration of BEC in low-dimensional systems for particles confined by
gravitational field or by a rotational container \cite{12}. Possible
experimental configurations for BEC in spin-polarized hydrogen in two
dimensions are currently being discussed \cite{7,12a}.

Recent numerical studies of the GP equation in three space dimensions
\cite{3,4,5,CHI,6,cst} in time-independent and time-dependent forms have
emphasized that extensive care in numerical integration is needed  to 
obtain  good convergence. 
With the viability of experimental detection of BEC in two space
dimensions \cite{7}, here we perform a numerical study of the
time-dependent and time-independent GP equation in two space dimensions
for an interacting Bose gas under the action of a harmonic oscillator trap
potential.  The interatomic interaction is taken to be both attractive and
repulsive in nature.

The nonlinear time-dependent and time-independent GP equations can be
compared with the corresponding two types of the linear Schr\"odinger
equation. The stationary states in both cases have a trivial time
dependence of the form $\Psi({\bf r},t)=\exp (-iEt/\hbar)\Psi({\bf r})$
where $E$ is the parametric energy and $t$ the time. As is well known the
time-independent form of these equations determines the stationary
function $\Psi({\bf r})$, as in the hydrogen-atom problem. 
The time-dependent Schr\"odinger equation can also be
directly solved to obtain the full time-dependent solution in the case of
the stationary problems, from which the trivial time dependence $\exp
(-iEt/\hbar)$ can be separated. 
In fact, the  time-dependent methods have  been successfully
used for the bound-state calculation in many areas of 
computational quantum chemistry \cite{mc}. 
This way of extracting the stationary
solution from the linear time-dependent Schr\"odinger equation continues
as a powerful technique in the case of nonlinear time-dependent GP
equations.

In this paper we solve the stationary BEC problem in two dimensions using
both the time-dependent and time-independent GP equations in the cases of
attractive and repulsive interatomic interactions and compare the two
types of solutions. The time-independent GP equation is solved by
integrating it with the Runge-Kutta rule complimented by the known
boundary conditions at origin and infinity \cite{koo}. The time-dependent
GP equation is solved by discretization and Gauss elimination method with
the Crank-Nicholson-type rule complimented again by the known boundary
conditions \cite{koo}. We find that both the time-dependent and
time-independent approaches lead to good convergence for the stationary
bound-state problem of the condensate. We also compare these solutions
with the Thomas-Fermi approximation in the case of repulsive interatomic
interaction.

In addition to obtaining the solution of the stationary problem the
time-dependent GP equation can be used to study the intrinsic
time-evolution problems with nontrivial time dependence and in this paper
the time-dependent approach is also used to study some evolution problems. 
Specifically, we study the effect of suddenly altering the trapping energy
on a preformed condensate. We find that in this case instead of executing
sinusoidal oscillations between the stable initial and final
configurations as in standard time-evolution problems governed by the
linear Schr\"odinger equation, the condensate executes oscillations around
the stable initial and final configurations with ever-growing amplitude.

In Sec. II we describe the time-dependent and time-independent forms of
the GP equation. In Sec. III we describe the numerical method in some
detail. In Sec. IV we report the numerical results and finally, in Sec. V
we give a summary of our investigation.

\section{Nonlinear Gross-Pitaevskii Equation}

At zero temperature, the time-dependent Bose-Einstein condensate wave
function $\Psi({\bf r},\tau)$ at position ${\bf r}$ and time $\tau $ may
be described by the self-consistent mean-field nonlinear GP equation
\cite{8}. In the presence of a magnetic trap this equation is written as
\begin{eqnarray}\label{a} \biggr[ -\frac{\hbar^2}{2m}\nabla^2
&+&\frac{1}{2}m\omega^2 r^2 +gN|\Psi({\bf r},\tau)|^2\nonumber \\
&-&i\hbar\frac{\partial
}{\partial \tau} \biggr]\Psi({\bf r},\tau)=0.   \end{eqnarray} Here $m$
is
the mass of a single bosonic atom, $N$ the number of atoms in the
condensate, $ m\omega^2 r^2/2$ the attractive harmonic-oscillator trap
potential, $\omega$ the oscillator frequency,
 and $g$ the strength of interatomic interaction. A positive $g$
corresponds to a repulsive interaction and a negative $g$ to an attractive
interaction. The normalization condition of the wave function is
\begin{equation}\label{4}
 \int_0^\infty d{\bf r} |\Psi({\bf r},t)|^2 = 1.  \end{equation}

For a stationary solution the time dependence of the wave function is
given by $\Psi({\bf r},\tau) = \exp(-i\mu \tau /\hbar) \Psi({\bf r})$
where $\mu$ is the chemical potential of the condensate. If we use this
form of the wave function in Eq. (\ref{a}), we obtain the following
stationary nonlinear time-independent GP equation \cite{8}: 
\begin{eqnarray}\label{1} \left[ -\frac{\hbar^2}{2m}\nabla^2
+\frac{1}{2}m\omega^2 r^2 +gN|\Psi ({\bf r})|^2-\mu \right]\Psi({\bf
r})=0.  \end{eqnarray} The time-dependent equation (\ref{a}) is equally
useful for obtaining a stationary solution with trivial time dependence as
well as for studying evolution processes with explicit time dependence.

Here we shall be interested in the spherically symmetric solution
$\Psi({\bf r},\tau)\equiv \varphi({ r},\tau)=\varphi(r)\exp(-i\mu
\tau/\hbar)$
to Eqs.  (\ref{a}) and (\ref{1}),
 which can be written, respectively, as \begin{eqnarray}\label{c} \biggr[
-\frac{\hbar^2}{2m}\frac{1}{r}\frac{\partial }{\partial r}r\frac{\partial
}{\partial r} &+& \frac{1}{2}m\omega^2 r^2 +gN|\varphi ({
r},t)|^2\nonumber \\
&-& i\hbar\frac{\partial}{\partial \tau}\biggr] \varphi(r,\tau)=0,
\end{eqnarray} \begin{eqnarray}\label{2} \biggr[
-\frac{\hbar^2}{2m}\frac{1}{r}\frac{d}{dr}r\frac{d}{dr}
&+&\frac{1}{2}m\omega^2 r^2 +gN|\varphi({ r})|^2
\nonumber \\&-&\mu \biggr]\varphi({
r})=0.
\end{eqnarray} The above limitation to the spherically symmetric solution
(in zero angular momentum state) reduces the GP equations in two physical
space dimensions to one-dimensional differential equations. We shall study
numerically these
 one-dimensional equations in this paper. 

As in Ref. \cite{6}, it is convenient to use dimensionless variables
defined by $x = r/a_{\mbox{ho}}$, and $t=\tau \omega/2, $ where
$a_{\mbox{ho}}\equiv \sqrt {\hbar/(m\omega)}$, $\alpha = \mu/(\hbar
\omega)$, $ \psi(x) = a_{\mbox{ho}}\sqrt{2mgN}\varphi(r)/\hbar$, and $
\psi(x,t) = a_{\mbox{ho}}\sqrt{2\pi}\varphi(r,\tau)$. In terms of these
 variables Eqs. (\ref{c}) and (\ref{2}) becomes, respectively,
\begin{eqnarray}\label{d}\left[ -\frac{1}{x}\frac{\partial }{\partial
x}x\frac{\partial}{\partial x} +x^2 +cn|\psi({x},t)|^2 -i\frac{\partial
}{\partial t} \right]\psi({ x},t)=0, \end{eqnarray}
\begin{eqnarray}\label{3} \left[ -\frac{1}{x}\frac{d}{dx}x\frac{d}{dx}
+x^2 +c|\psi({x})|^2-2\alpha \right]\psi({ x})=0.  \end{eqnarray} where
$n\equiv mgN/(\pi \hbar^2)$ is the reduced number of particles and $c=\pm
1$ carries the sign of $g$: $c=1$ corresponds to a repulsive
 interaction and $c=-1$ corresponds to an attractive interaction.
The normalization condition (\ref{4}) of the wave
functions become \begin{equation}\label{5} 1= \int_0 ^\infty |\psi(x,t)|
^2 xdx
 = \frac{1}{n}\int_0 ^\infty |\psi(x)|^2 xdx.  \end{equation} We shall be
using these two slightly different normalizations of the time-dependent
and time-independent wave functions for future numerical convenience.

An interesting property of the condensate wave function is its mean-square
radius defined by \begin{equation}\label{7} \langle x^2 \rangle= \int_0
^\infty x^2 |\psi(x,t)| ^2 xdx
 = \frac{1}{n}\int_0 ^\infty x^2 |\psi(x)|^2 xdx.  \end{equation}

\section{Numerical Method} 
\subsection{Boundary Condition}

Both in time-dependent and time-independent approaches we need the
boundary conditions of the wave function as $x \to 0$ and $\infty$. For a
confined condensate, for a sufficiently large $x$, $\psi(x)$ must vanish
asymptotically. Hence the nonlinear term proportional to $|\psi(x)|^3$ can
eventually be neglected in the GP equation for large $x$ and Eq. (\ref{3}) 
becomes \begin{eqnarray}\label{8} \left[
-\frac{1}{x}\frac{d}{dx}x\frac{d}{dx} +x^2 -2\alpha \right]\psi({ x})=0. 
\end{eqnarray} This is the equation for the oscillator in two space
dimensions in the spherically symmetric state with solutions for $\alpha =
1, 3, 5,...$ etc. A general classification of all the states of such an
oscillator is well under control \cite{ho}. In the present BEC problem,
Eq. (\ref{8}) determines only the asymptotic behavior.  If we consider Eq.
(\ref{8}) as a mathematical equation valid for all $\alpha$ and large $x$,
the asymptotic form of the physically acceptable solution is given by
\begin{equation}\label{9} \lim_{x \to \infty} \psi(x)= N_C\exp
\left[-\frac{x^2}{2}+(\alpha - 1)\ln x \right], \end{equation} where
$N_C$ is a normalization constant. Equation (\ref{9}) leads to the
following asymptotic log-derivative \begin{equation}\label{10} \lim_{x \to
\infty} \frac {\psi'(x)}{\psi{(x)}}= \left[-x+ \frac{\alpha-1}{x}\right],
\end{equation} which is independent of the constant $N_C$ and where the
prime denotes derivative with respect to $x$. 

Next we consider Eq. (\ref{3}) as $x\to 0$. The nonlinear term approaches
a constant in this limit because of the regularity of the wave function at
$x=0$. Then one has the following usual conditions \begin{equation}
\psi(0)={ \mbox{constant}}, \hskip 1cm \psi'(0)=0 , \label{11}
\end{equation} as in the case of the harmonic oscillator problem in two
space dimensions \cite{ho}.  Both the small- and large-$x$ behaviors of
the wave function will be necessary for a numerical solution of the GP
equation in time-dependent and time-independent forms.

\subsection{Time-Dependent Approach: Evolution and Stationary Problems}

First we describe the numerical method for solving the time-dependent
equation (\ref{d}). For a numerical solution it is convenient to make the
substitution $\psi(x,t)\equiv \phi(x,t)/x$ in this equation, when this
equation becomes \begin{eqnarray}\label{e}\biggr[
-\frac{\partial^2}{\partial x^2} &+& \frac{1}{x} \frac{\partial}{\partial
x}
-\frac{1}{x^2} +x^2 +cn\frac{|\phi({x},t)|^2
}{x^2}\nonumber \\ &-& i\frac{\partial}{\partial t} \biggr]\phi({ x},t)=0.
\end{eqnarray}
A convenient way to solve Eq. (\ref{e}) numerically is to discretize it in
both space and time and reduce it to a set of algebraic equations which
could then be solved by using the known asymptotic boundary conditions. 
We discretize this equation by using a space step $h$ and time step
$\Delta $ with a finite difference scheme using the unknown $\phi^k_j$
which will be approximation of the exact solution $\phi(x_j,t_k)$ where
$x_j= j h$ and $t_k=k\Delta $.  As Eq. (\ref{e}) involves both time and
space variables it can be discretized in more than one way.  The time
derivative in Eq. (\ref{e}) involves the wave function at times $t_k$ and
$t_k+\Delta$. As $\Delta$ is small, the time-independent operations in
this equation can be discretized by using the wave-function components at
time $t_k$ or $t_{k+1}\equiv t_k+\Delta$.  If one uses the wave-function
components at time $t_k$, Eq. (\ref{e}) is discretized as \begin{eqnarray}
\frac{i(\phi_j^{k+1}-\phi_j^{k}) }{\Delta } &=& -\frac{1}{h ^2}
\left[\phi^{k}_{j+1}-2 \phi^{k}_{j}+\phi^{k}_{j-1}\right]\nonumber \\ 
&+&\frac{1}{2x_j h} \left[\phi^{k}_{j+1}- \phi^{k}_{j-1}\right]\nonumber
\\
&+&\left[x_j^2-\frac{1}{x_j^2}+cn\frac{|\phi_j^{k}|^2}{x_j^2}\right]
\phi_j^k.\label{fx} \end{eqnarray} This is an explicit differencing
scheme, since, given $\phi$ at $t_k$ it is straightforward to solve for
$\phi$ at $t_{k+1}$ \cite{koo}. One should start with an approximately
known solution at $t_k$ and propagate it in time until a converged
solution is reached.    We  confirm in
our study that this simple scheme leads to slow convergence and large
unphysical oscillations in the solution. 

One can express the derivatives on the right-hand-side of Eq. (\ref{fx}) 
in terms of the variables at time $t_{k+1}$ \cite{koo}. Then the unknown
$\phi_j^{k+1}$ appears on both sides of the equation and one has an
implicit scheme. We find that the implicit scheme improves substantially
the numerical accuracy and convergence rate. However, we find after some
experimentation that if the right-hand-side of Eq. (\ref{fx}) is averaged
over times $t_k$ and $t_{k+1}$ one has the best convergence. This is a
semi-implicit scheme based on the Crank-Nicholson scheme for
discretization \cite{koo1}. We use the following rule  to discretize the
partial differential equation (\ref{e})
\cite{koo,koo1} \begin{eqnarray} \frac{i(\phi_j^{k+1}-\phi_j^{k}) }{\Delta
} &=& -\frac{1}{2h ^2}\biggr[(\phi^{k+1}_{j+1}-2
\phi^{k+1}_{j}+\phi^{k+1}_{j-1}) \nonumber \\ &+& (\phi^{k}_{j+1}-2
\phi^{k}_{j}+\phi^{k}_{j-1})\biggr]\nonumber \\ &+&\frac{1}{4x_j h}
\left[(\phi^{k+1}_{j+1}- \phi^{k+1}_{j-1})+(\phi^{k}_{j+1}-
\phi^{k}_{j-1})\right]\nonumber \\
&+&\frac{1}{2}\left[x_j^2-\frac{1}{x_j^2}+cn\frac{|\phi_j^{k}|^2}{x_j^2}\right]
(\phi_j^{k+1}+\phi_j^k).\nonumber  \\ \label{f} \end{eqnarray} Similar
discretization
rule has been used for the solution of the GP equation in three space
dimensions \cite{4}.  The first and second space derivatives of the wave
function as well as the
 the wave function itself have been approximated by the average over their
values at the initial time $t_k$ and the final time $t_{k+1}$. This
procedure leads to accurate and stable numerical results. Considering that
the wave function is known at time $t_k$, Eq. (\ref{f}) is an equation in
three unknowns $-$ $\phi_{j+1}^{k+1},\phi_j^{k+1}$ and $\phi_{j-1}^{k+1}$. 
In a lattice of $N$ points Eq. (\ref{f}) represents a tridiagonal set for
$j=2,3,...,(N-1)$. This set has a unique solution if the wave functions at
the two end points $\phi_{1}^{k+1}$ and $\phi_{N}^{k+1}$ are known. In the
present problem these values at the end points are provided by the known
asymptotic conditions. The tridiagonal set of equations is solved by the
Gauss elimination method and back substitution \cite{koo} using a typical
space step $h = 0.0001$ and time step $\Delta = 0.03$. Although, the
iterative method should work for any value of $\Delta$, we found the
convergence to be faster with this value of $\Delta$ and we used this
value throughout the present investigation.

The time-dependent method could be used to study stationary as well as
time-evolution problems. First we consider the stationary problem. For the
ground and the first excited states of the condensate we start with the
following analytically known wave functions of the harmonic oscillator
problem (\ref{8}) \cite{ho}:  \begin{equation} \phi(x)= x\psi(x)=\sqrt 2
x\exp (-x^2/2), \end{equation} \begin{equation} \phi(x)= x\psi(x)=\sqrt 2
x(1-x^2)\exp (-x^2/2), \end{equation} respectively, at an initial time
$t=0$. We then repeatedly propagate these solutions in time using the
Crank-Nicholson-type algorithm (\ref{f}). The boundary condition
(\ref{11}), that $\phi(0)=0$, is implemented at each time step \cite{koo}. 
Also, the solution at each time step will satisfy the asymptotic condition
(\ref{9}). Starting with $cn=0$, at each time step we increase or decrease
the nonlinear constant $cn$ by an amount $\Delta_1$ typically around
0.01.  This procedure is continued until the desired final value of
$cn$ is reached.  Then the final solution is iterated several times
(between 10 to 40 times) to obtain a stable converged result.
The resulting solution is the ground state of the
condensate corresponding to the specific nonlinear constant $cn$. We found
the convergence to be fast for small $|cn|$.  However, the final
convergence
of the scheme breaks down if $|cn|$ is too large. In practice these
difficulties start for $cn>20$ for the ground state for a positive $c$
(repulsive interaction) in a computational analysis in double precision. 
For an attractive interaction there is no such problem as the GP equation
does not sustain a large nonlinearity $|cn|$ as we comment in detail in
the next section. 

As the time dependence of these stationary states is trivial $-$
$\psi(x,t)= \psi(x) \exp(-i2\alpha t)$ $-$ the chemical potential $\alpha$
can be obtained from the propagation of the converged ground-state
solution at two successive times, e.g., $\psi(x,t_k)$ and
$\psi(x,t_{k+1})$.  From the numerically obtained ratio
$\psi(x,t_k)/\psi(x,t_{k+1})= \exp(i2\alpha \Delta)$ $\alpha$ can be
obtained as the time step $\Delta $ is known.

The time-dependent method could also be used to study evolution problems. 
One such evolution problem describes the fate of the condensate if the
trap potential is removed or altered suddenly after the formation of the
condensate. As a stable condensate is formed under the action of the trap
potential, after a sudden change in the trap potential, the condensate
will gradually modify with time. To study the time evolution of a
condensate wave function as the trap is removed or altered suddenly, we
have to start the time evolution of the known precalculated wave function
of the condensate with the initial trap potential and allow it to evolve
in time using the time-dependent GP equation with the full nonlinearity
but with the altered trap potential, which could be zero.

\subsection{Time-Independent Approach}

The time-independent GP equation (\ref{3}) has the following structure
\begin{eqnarray} y'= G(x, \psi(y)), \end{eqnarray} with $y=x\psi'$, where
the prime denotes the $x$ derivative. With this realization, a numerical
integration of Eq.  (\ref{3}) can be implemented using the following
four-point Runge-Kutta rule \cite{koo,w} in steps of $h$ from $x_j$ to
$x_{j+1}$ \begin{eqnarray}\label{g} \psi_{j+2}&
=&\psi_{j+1}+h
\psi'_{j+1},\\ x_{j+1}\psi'_{j+1}&=&x_j\psi'_j +
\frac{1}{6}(s_0+2s_1+2s_2+s_3), \end{eqnarray}
 where \begin{eqnarray}\label{h} s_0&=&h G(x_j,\psi_j),\\ s_1&=& h
G\left[x_j+\frac{h}{2},\psi_j+\frac{h (x_j
\psi'_j+{s_0}/{2})}{2(x_j+h/2)}\right],\\ s_2&=& h
G\left[x_j+\frac{h}{2},\psi_i+\frac{h(x_j
\psi'_j+{s_1}/{2})}{2(x_j+h/2)}\right],\\ s_3&=& h
G\left[x_j+{h},\psi_j+\frac{h (x_j \psi'_j+{s_2})}{(x_j+ h)}\right]. 
\end{eqnarray}
 Equation (\ref{3}) is integrated numerically for a given $\alpha$ using
this algorithm starting at the origin ($x=0$) with the initial boundary
condition (\ref{11}) with a trial $\psi(0)$ and a typical space step $h =
$ 0.0001. The integration is propagated to $x = x_{\mbox {max}}$, where
the asymptotic condition (\ref{10}) is valid. The agreement between the
numerically calculated log-derivative of the wave function and the
theoretical result (\ref{10})  was enforced to five significant figures. 
The maximum value of $x$, up to which we needed to integrate (\ref{3})
numerically for obtaining this precision, is $x_{\mbox {max}}= 5 $. If for
a trial $\psi(0)$, the agreement of the log-derivative can not be
obtained, a new value of $\psi(0)$ is to be chosen.  The proper choice of
$\psi(0)$ was implemented by the secant method. Even with this method,
sometimes it is difficult to obtain the proper value of $\psi(0)$ for a
given $\alpha$.  Unless the initial guess is ``right"  and one is
sufficiently near the desired solution, the method could fail, specially,
for large $|cn|$ and lead numerically to either the trivial solution
$\psi(x) = 0$ or an exponentially divergent nonnormalizable solution in
the asymptotic region.

\section{Numerical Result}

\subsection{Stationary Problem}

First we consider the ground-state solution of Eq. (\ref{3}) for different
$\alpha $ in cases of both attractive and repulsive interactions using the
time-independent method.  In the presence of the nonlinearity, for
attractive (repulsive) interatomic interaction, the solutions of the GP
equation for the ground state appear for values of chemical potential
$\alpha <1$ ($\alpha > 1$).  The relevant parameters for the solutions
$-$ the wave-function at the origin $\psi(0)$, reduced number
$n$,
and mean-square radii $\langle x^2 \rangle$ $-$  are listed in Table I. 
The numerical integration was performed
up to $x_{\mbox{max}}=5$ with $h=0.0001$
where the asymptotic boundary condition (\ref{10})
is implemented. 

Using the known tabulated values of $n$ in each case we also solved the
time-dependent GP equation and the wave functions and energies so
calculated agree well with the respective quantities calculated with the
time-independent approach.  The solutions were obtained using space step
$h=0.0001$, time step $\Delta=0.03$ and the parameter $\Delta_1\approx
0.01$.  
The largest value of $x$ used in discretization (\ref{f}) is
$x_{\mbox{max}}=10$.
The wave functions for different values of $\alpha$ (and $n$) for
the attractive and repulsive interparticle interactions for the cases
shown in Table I are exhibited in Figs. 1(a) and 1(b), respectively, where
we plot $\psi(x)$ versus $x$ using the time-dependent and time-independent
approaches. The curves in Figs.  1(a) and 1(b) appear in the same order
as the rows in Table I and it is easy to identify the corresponding values
of $\alpha$ from the values of $\psi(0)$ of each curve. From Figs. 1(a) 
and (b) we find that the nature of the wave function for these two cases
are quite different. However, the wave functions calculated with
time-dependent and time-independent approaches agree reasonably with each
other. 

\vskip -.5cm
\vskip -3.8cm
\postscript{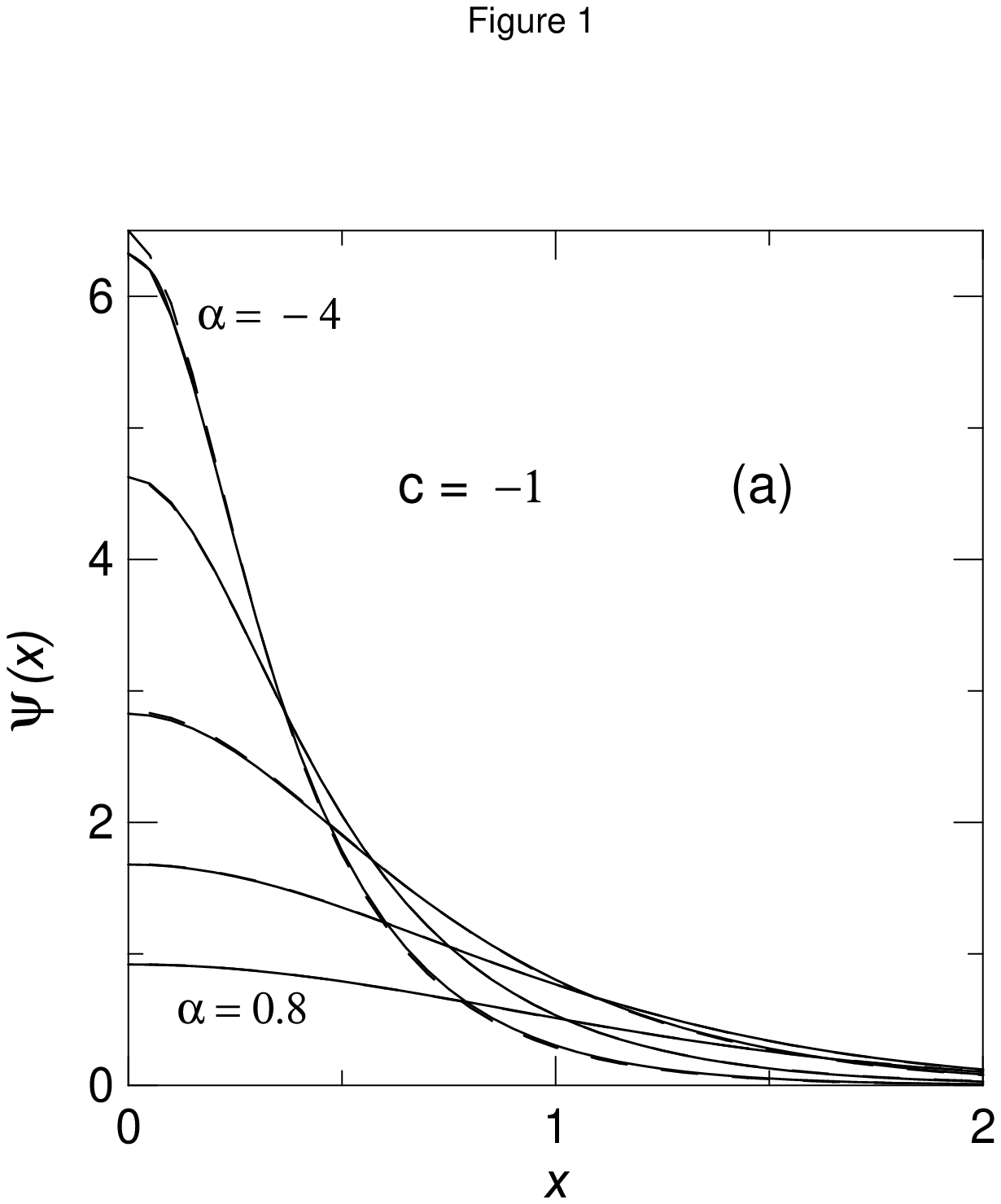}{1.0}    
\vskip -2.1cm
\vskip -3.8cm
\postscript{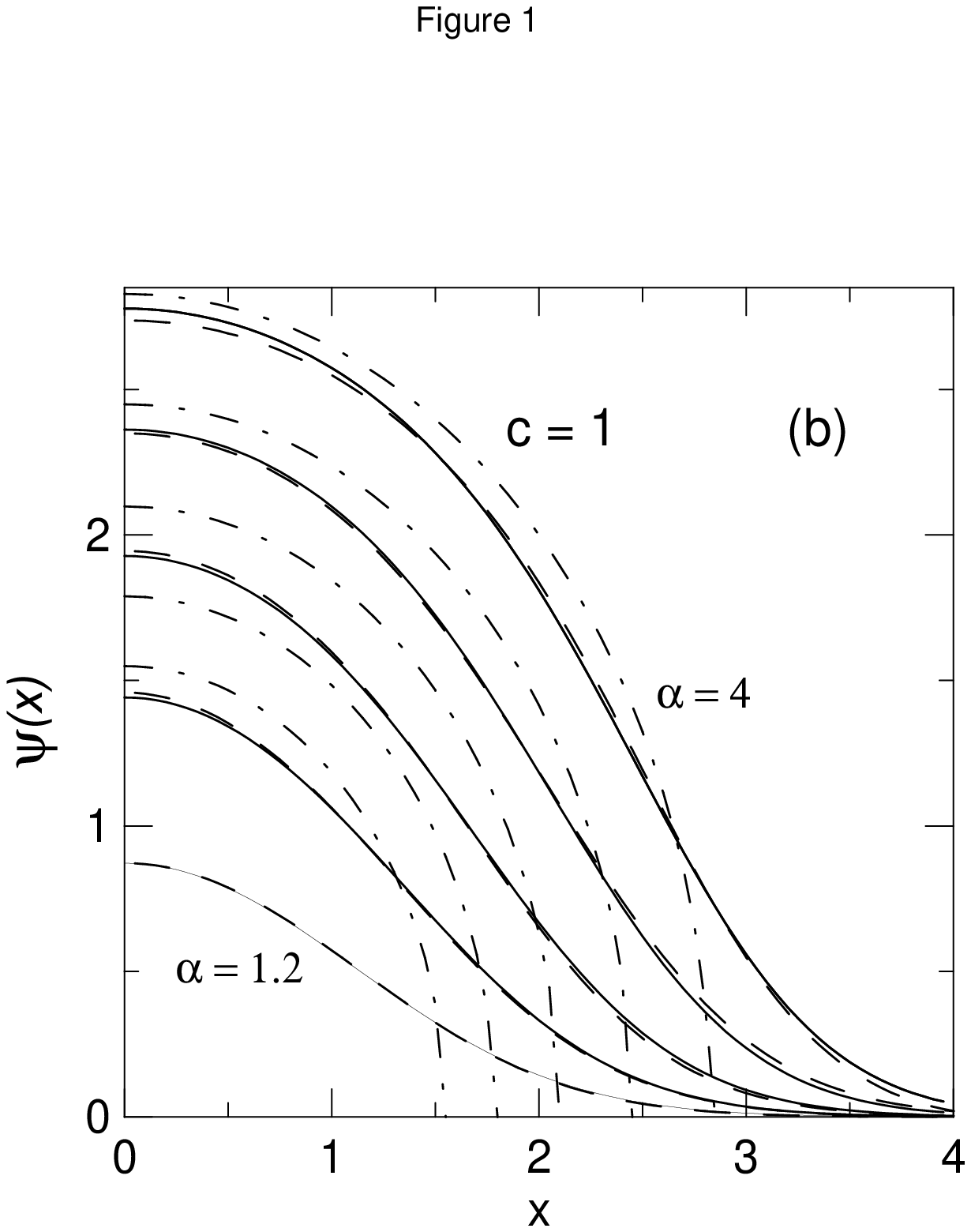}{1.0}    
\vskip -2.1cm

\vskip .4cm

{ {\bf Fig. 1.}   Ground-state condensate wave function $\psi(x)$ versus $x$ for (a) 
attractive and (b) repulsive interparticle interactions using the
time-dependent (dashed line) and time-independent (full line) approaches.
The parameters for these cases are given in Table I. In the time-dependent
method we used time step $\Delta=0.03, \Delta_1 = 0.01$, space step
$h=0.0001$ and $x_{\mbox{max}} = 8$, in the time-independent method we
used space step $h=0.0001$ and $x_{\mbox{max}} = 5.$ In the case of the
repulsive interparticle interaction we also show the solution (\ref{tf})
corresponding to the Thomas-Fermi approximation (dashed-dotted line).  The
curves appear in same order as in Table I. with the lowermost curve
corresponding to the first row.}

In the absence of previous solutions of this problem we compare the
stationary solutions in the repulsive case ($c=1$) with those obtained via
a well-known approximation, e.g., the Thomas-Fermi approximation. In this
approximation the kinetic energy term in Eq.  (\ref{3}) is neglected and
one has the following simple approximate solution
\begin{equation}\label{tf} \psi(x) = \sqrt{2\alpha -x^2}, \end{equation}
for $x^2\le 2\alpha$ and zero otherwise. In Fig. 1 (b) we also plot the
Thomas-Fermi approximation (\ref{tf}).  We find that as expected, for a
large condensate, this approximation is a reasonable approximation.
However, it turns out to be a bad approximation for a small condensate.

\vskip .15cm

{Table I: Parameters for the numerical solution of the GP equation
(\ref{3}) for $c=\pm 1$ for the ground state wave function. The first four
columns refer to the attractive interaction $c=-1$ and the last four
columns refer to the repulsive interaction $c=1$.  } \vskip .2cm

{\begin{center}{\begin{tabular} {|c|c|c|c|c|c|c|c|} \hline $\alpha$ &
$\psi(0)$ & $n$ & $\langle x^2\rangle$ & $\alpha$ & $\psi(0)$ & $n$ &
$\langle x^2\rangle$ \\ \hline 1.0 & 0 & 0 & 0 & 1.0 &0 &0 & 0 \\ 0.8 &
0.9185 & 0.3663 & 0.9030 &1.2 & 0.8719& 0.4353&1.1027 \\ 0.4 & 1.6795 &
0.9147 & 0.7297 & 1.6 &1.4415 &1.5276 &1.3219\\
   $-$0.4 & 2.8255 & 1.4798 & 0.4757 & 2.2 &1.9276 &3.7509 &1.6741\\
$-$2.0 & 4.6249 & 1.7695 & 0.2400 & 3.0 & 2.3626&7.8377 &2.1679 \\
$-$4.0&6.3252&1.8319&0.1385 & 4.0 & 2.7786& 14.7609 &2.8041 \\ \hline
\end{tabular}}\end{center}} \vskip .15cm

It is appropriate to comment on the numerical accuracy of the present
time-dependent and independent methods, which seems to be limited
typically by the
difference between the time-dependent and independent solutions in Fig. 1.
When the solution can be obtained numerically, as in the cases shown in
Table I, the time-independent method can yield very accurate results. This
accuracy can be increased by controlling the space step $h$ and
$x_{\mbox{max}}$. This is not so in the case of the time-dependent method,
where the numerical result exhibits small periodic oscillation
after 
iteration specially for large values of $|cn|$ 
which we detail below.
 
The numerical solution of the time-dependent method is independent of the
space step $h$ provided that
a typical value around $h=0.0001$ is employed as in the present study. No
visible difference in the solution is found if $h$ is increased by a
factor of 2 or 3. However, the solution is more sensitive to the number 
of time iterations, specially, for a large value of $|cn|$,  for a fixed  
integration time step $\Delta$ or the step $\Delta_1$ by which the
nonlinear constant in the GP equation is increased at each time step until
the final value of $cn$ is reached. We show this variation in Figs. 2 (a)
and (b) where  we plot $|\psi(x,T)|$ as a function of reduced
time
$T\equiv t/0.03$ for $x =0$ and 2 for different choices of $\Delta$ and
$\Delta_1$ in the repulsive case for $n= 3.7509$ and $\alpha = 2.2$
corresponding to the fourth row of Table I. The zero of reduced time $T$
is made to coincide with the time step $t_k$ at which the full nonlinear
constant $cn$ is obtained for the first time during iteration. This choice
of time will allow us to compare the fluctuations of the solution during
the time propagation of the full GP equation.
 In Fig. 2(a) we present our results for $\Delta = 0.03$ and for
$\Delta_1=$  0.018754 and 0.0046886. In Fig. 2(b) we present our
results for $\Delta_1 = 0.0046886$ and for $\Delta=  0.03,$ and 0.05. 
From Figs. 2 (a) and (b) we find that there is numerical oscillation
of the solution with time in this approach which is independent of small
variations of $\Delta$ near 0.03 and $\Delta_1$ around  0.01. These
oscillations determine the numerical error of the time-dependent
approach and  become larger when we employ a $\Delta$ very different
from 0.03, or $\Delta_1$ very different from 0.01. The oscillations can
really be large if an improper value of step $\Delta$ or $\Delta_1$ is
choosen as can be seen from Fig, 2(b) for $\Delta=0.05$.  The results
remain stable if we reduce these steps up to $\Delta\sim 0.01$ and
$\Delta_1 \sim
0.003$. 
For very small
$\Delta_1$ and $\Delta$ accumulative errors also increase. 
This accumulative numerical error increases as the number of iterations is
very large (several thousands) and a large number of iterations is needed 
to cover a given time interval with a small time step $\Delta$. 

{Table II: Amplitude of oscillation $A(x,T)$ (in units of 0.01)
of  $|\psi(x,T)|$ at different times $T$
for $x=0$ and 2 calculated with $\Delta=0.03$ and $\Delta_1= 0.0046886$
in the repulsive case for $cn=3.7509$. The average value of converged 
$|\psi(0,T)| =1.9310$ and $|\psi(2,T)|=0.6667.$ 
} 

{\begin{center}{\begin{tabular} {|c|c|c|c|c|c|c|c|} 
\hline   $T=$ &0
 & 167  &  294 &  406 &  533 &  645 &
 791 \\ 
$|A(0,T)|$ & 1.54 & 1.27 & 3.32 & 2.93 &4.13
&4.49 &
7.35 \\
\hline   $T=$ &0
 & 162  &  291 &  400 &  536 &  637 &
 789 \\ 
$|A(2,T)|$ & 0.77 & 0.95 & 1.13 & 1.33 &1.43
&2.01 &
2.77\\
\hline
\end{tabular}}\end{center}} 
 
\vskip -.3cm
\vskip -3.8cm
\postscript{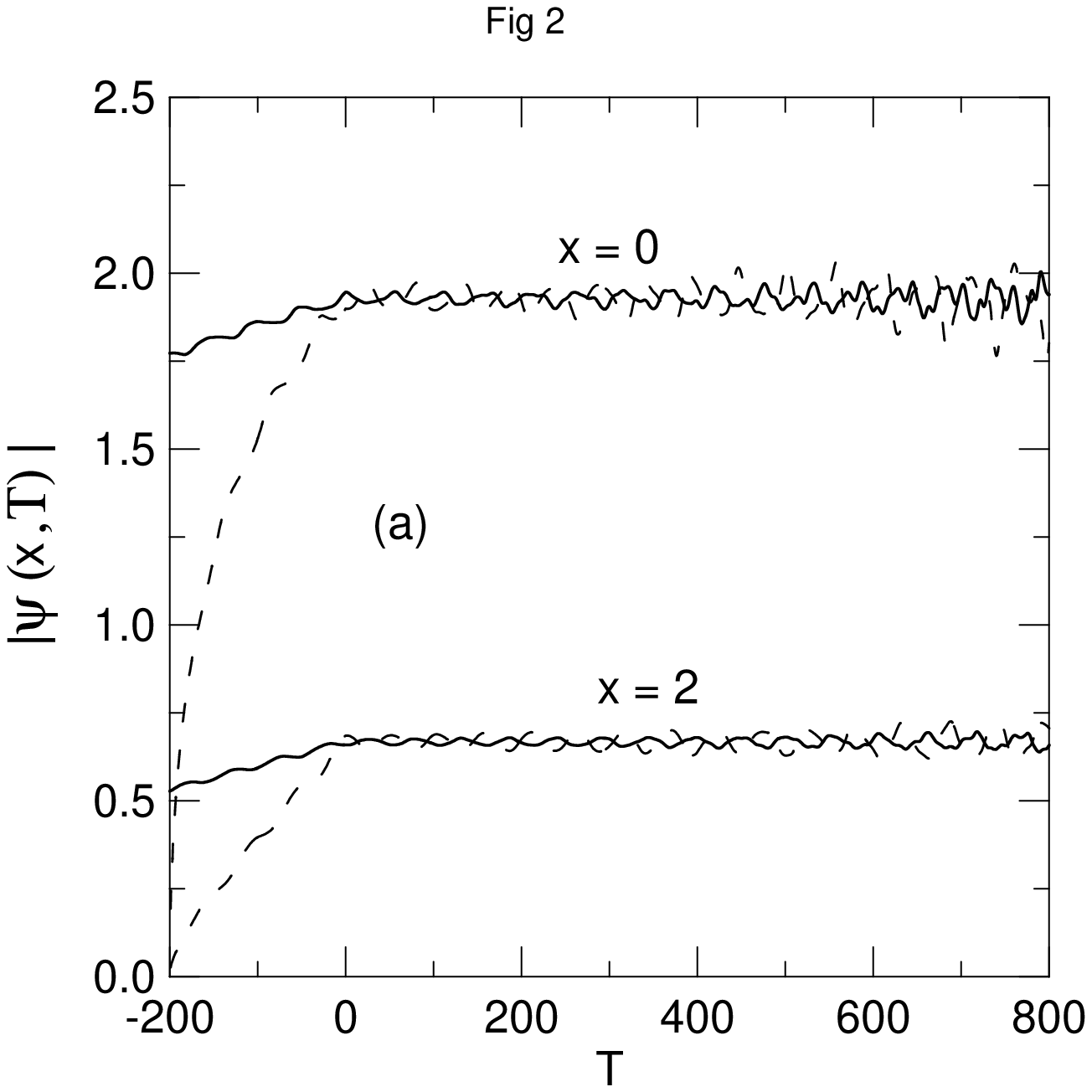}{1.0}    
\vskip -2.1cm
\vskip -3.8cm
\postscript{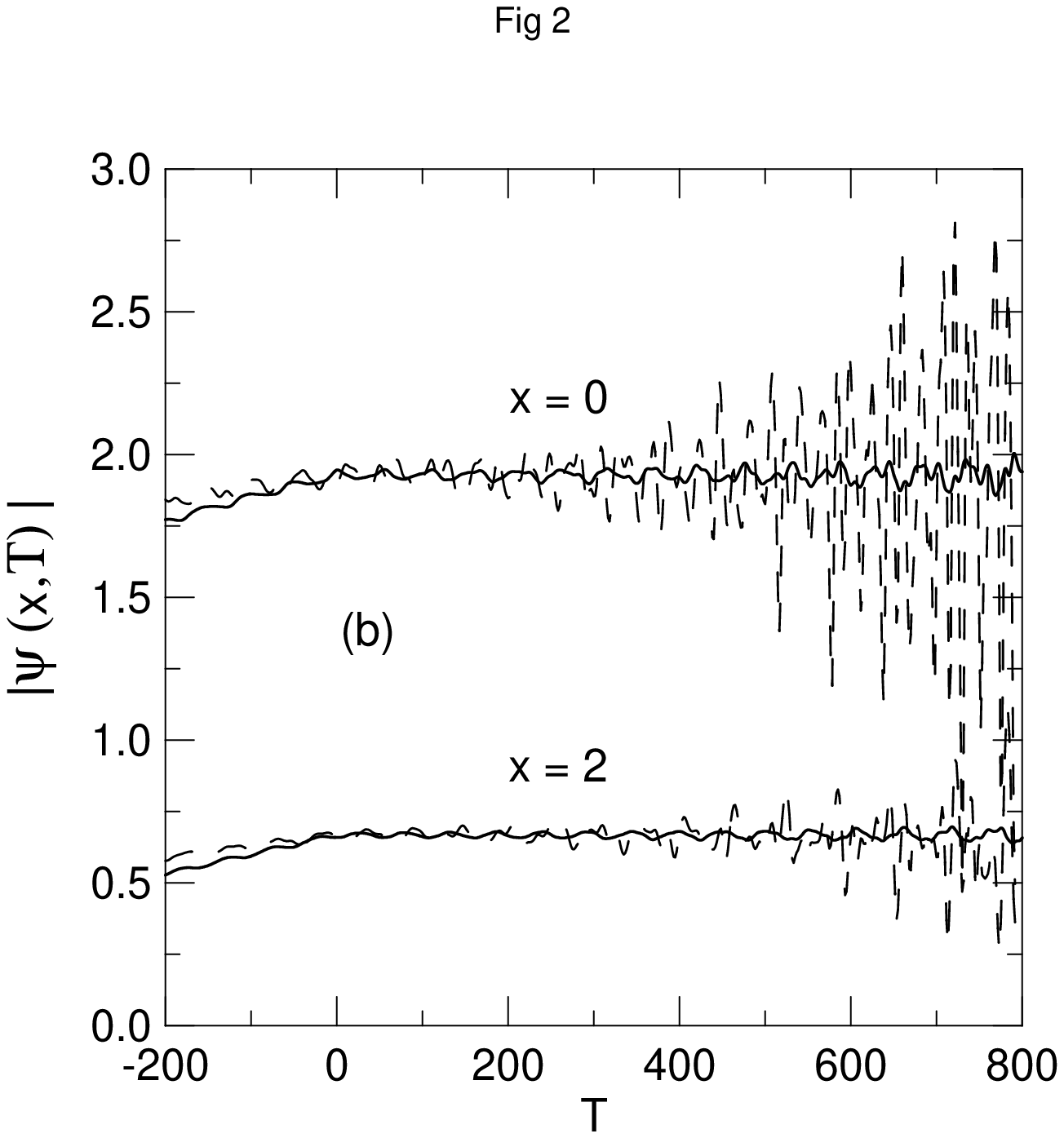}{1.0}    
\vskip -2.cm

{ {\bf Fig. 2.}    Ground-state condensate wave function $|\psi(x,T)|$
versus reduced time
$T\equiv t/0.03$ for $x=0$ and 2 in the repulsive case for the nonlinear
constant $cn=3.7509$ for (a) $\Delta =0.03$, and $\Delta_1= 
0.0046886$ (full line), and 0.018754 (dashed line)
and (b)  $\Delta_1 = 0.0046886$ and for $\Delta=$ 
0.03 (full  line), and 0.05 (dashed line).  The zero of $T$ is taken to be
the time at which the full nonlinearity is achieved for the first time. }

We show a
quantitative account of the above oscillation in Table II where we plot
the
maximum 
error 
in $|\psi(x,T)|$ (amplitude of oscillation of $|\psi(x,T)|$)
for $x=0 $ and 2 at different times calculated with steps  $\Delta=0.03$
and $\Delta_1= 0.0046886$.
We find that the error increases slowly, but not necessarily
monotonically, with time. 
The average value of the converged $|\psi(0,T)|
$ is 1.9310 and that for $|\psi(2,T)|$ is $0.6667.$ The maximum deviations
from these
values as shown in Table II do not occur at the same values of $T$. We
find from Table II that for small $T(\sim 0)$ the maximum average
error
in  $|\psi(x,T)|$ is about $1\%$. For $T \sim 800 $ this maximum average 
error  could be as high as $4\%$. As these errors are oscillating with 
time,
at a given $T$ this error could be smaller or
even zero. Considering that we are dealing with nonlinear equations these
errors are well within the acceptable limits.
 The errors shown in Table II
would also be the typical errors in time-evolution problems which we study
in the next subsection.

For repulsive interaction, it was increasingly difficult to find the
solution of the GP equation using both time-dependent and time-independent
methods for larger nonlinearity than those reported in Figs. 1 (a) and
(b).  The inputs of the time-independent method are $\alpha$ and an
appropriate $\psi(0)$. In this method it became difficult (or impossible) 
to find the appropriate $\psi(0)$ and find a solution for large $cn
(>20)$.  For large nonlinearity the secant method led to radially excited
state for the appropriate $\psi(0)$. In the time-dependent method the only
input is the value of $cn$. For a large $cn$ in the repulsive case, the
numerically obtained solution for the wave function shows many
oscillations and is clearly unacceptable physically.  A
Crank-Nicholson-type approach was also used to solve the GP equation in
three space dimensions \cite{4}. The numerical instability also set a
limit in that investigation in finding stationary ground-state solution
for large values of nonlinearity.

For attractive interparticle interaction, the wave function is more
sharply peaked at $x=0$ than in the case of the repulsive interparticle
interaction and one has a smaller reduced number $n$ and mean square
radius
$\langle x^2 \rangle$.  In this case we find from Table I that with a
reduction of the chemical potential $\alpha$, the reduced number $n$
increases slowly and the mean square radius $\langle x^2 \rangle$
decreases rapidly, so that the density of the condensate $\rho \equiv
n/\langle x^2 \rangle$ tends to diverge as $n$ tends to a maximum value
$n_{\mbox{max}}$. The increase in density lowers the interaction energy.
The kinetic energy of the system is responsible for the stabilization. As
the central density increases further for stronger attractive
interparticle interaction, kinetic energy can no longer maintain
equilibrium of the system and the system collapses.  Consequently, for
$n>n_{\mbox{max}}$, there is no stable solution of the GP equation. 
Numerically, from a plot of $n$ versus $1/\rho$ we find this maximum
number consistent with $\rho^{-1}=0$ to be \begin{equation}\label{12}
n_{\mbox{max}}\equiv \eta N_{\mbox{max}} \approx 1.88.  \end{equation}
There is no such limit on $n$ in the repulsive case. In that case with the
increase of the chemical potential $\alpha$ the condensate increases in
size as the number of particles in the condensate increases.  These
behaviors of the Bose-Einstein condensate in two dimensions were also
noted in three dimensions \cite{11,13}.  However, in three dimensions the
corresponding maximum value was $n_{\mbox{max}}\equiv
4N_{\mbox{max}}|a|/a_{\mbox{ho}} \approx 2.30$ \cite{11}.

Both the time-dependent and time-independent approaches are equally
applicable for spherically-symmetric radially excited states. For the
first excited state, with one node in the wave function, we verified that
the convergence was as good as in the ground-state case reported here. 
However, it is a routine study and we do not report the results here.

\subsection{Evolution Problem}
\vskip -4.5cm
\postscript{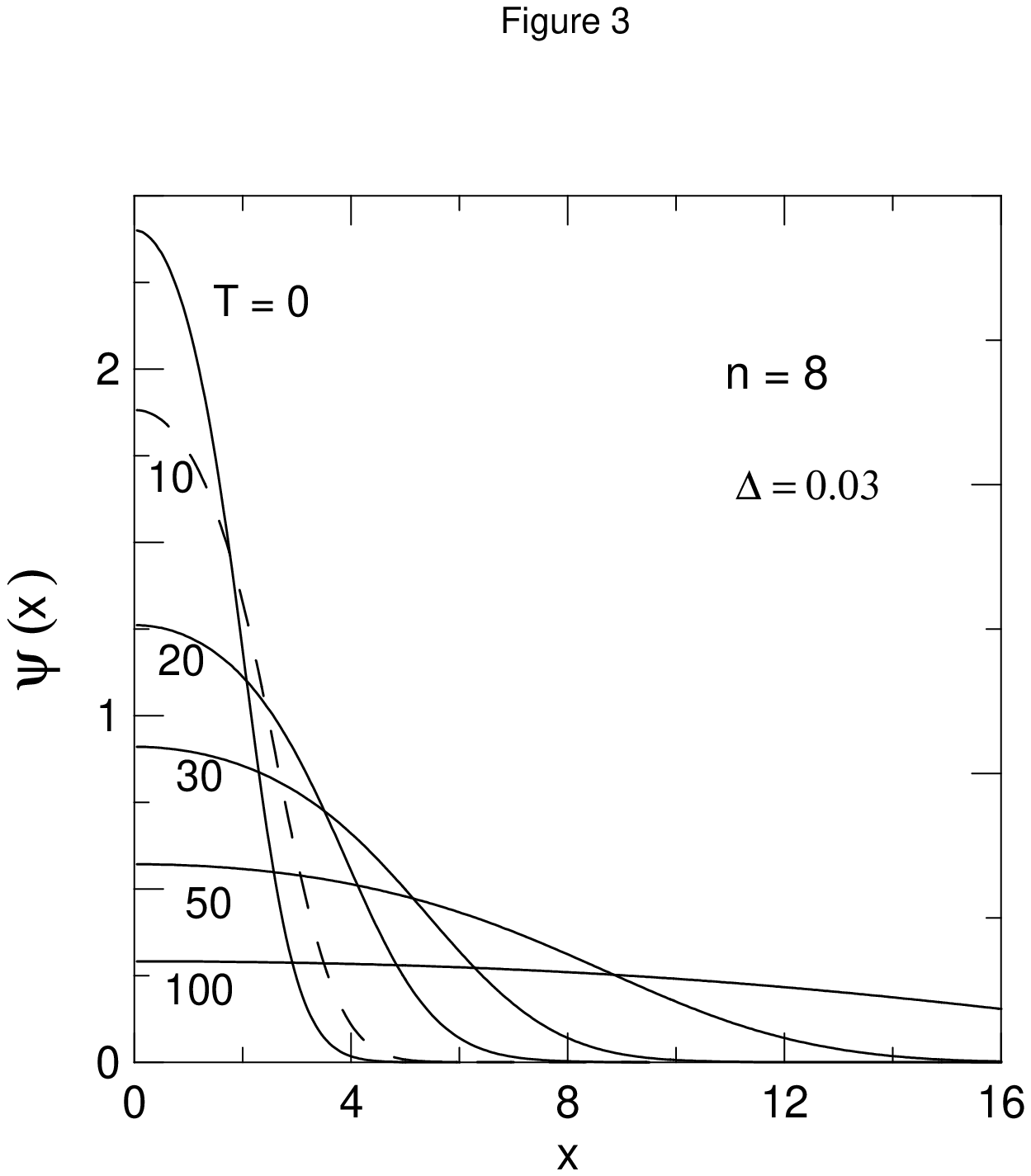}{1.0}    
\vskip -2.1cm

\vskip .4cm

{ {\bf Fig. 3.} 
 Condensate wave function $\psi(x)$ in the repulsive case at different
times $T=t/0.03$ for an expanding condensate after the trap is removed
suddenly at $T=0$.  The initial condensate has $n=8$, and the time
evolution is performed using time step $\Delta=0.03, \Delta_1 =0.01$,
space step $h=0.0001$ and $x_{\mbox{max}} = 15$. }

\vskip -3.8cm
\postscript{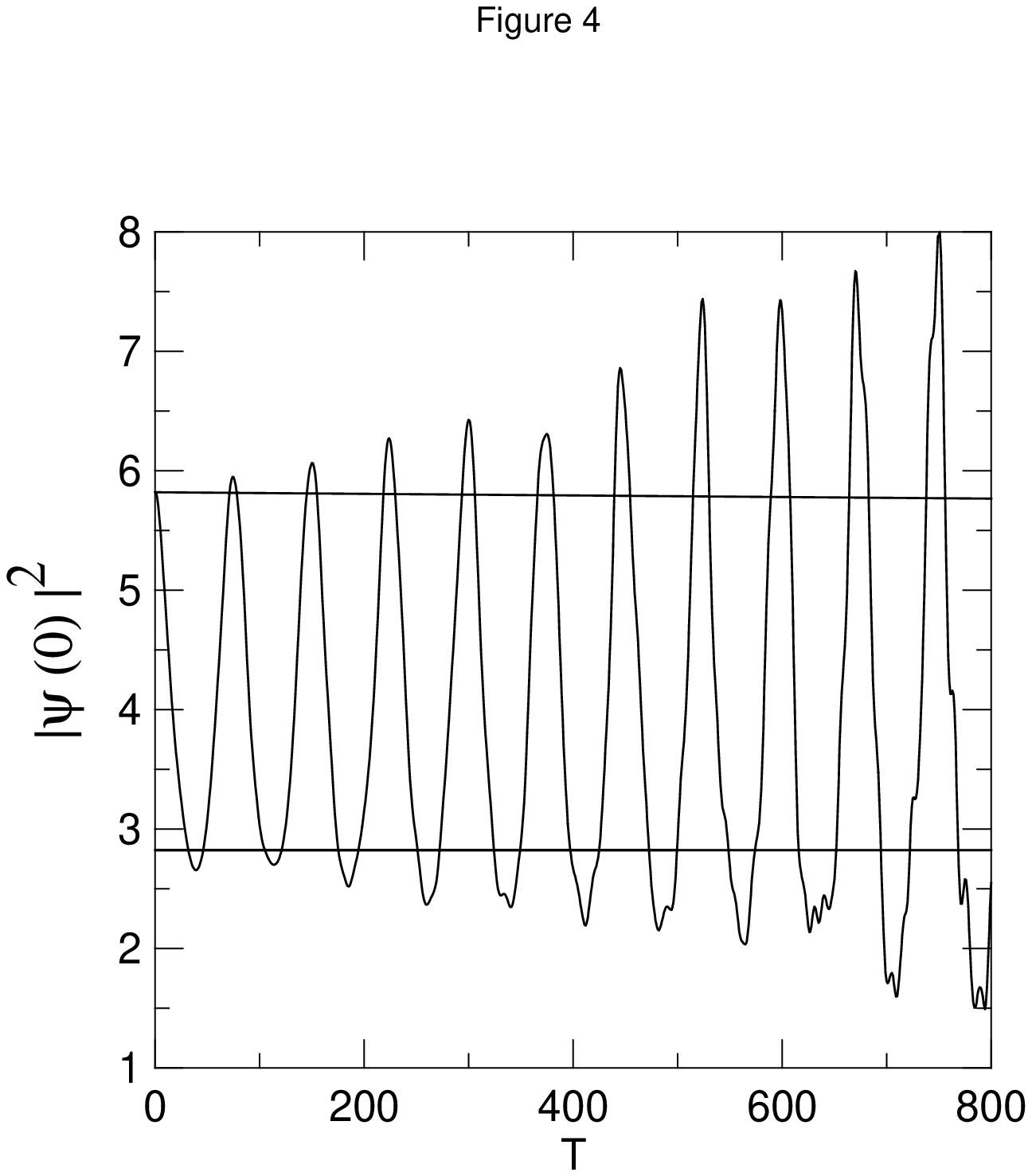}{1.0}    
\vskip -2.1cm

\vskip .4cm

{ {\bf Fig. 4.} The central probability density $|\psi(0)|^2$ in the repulsive case at
different times $T=t/0.03$ for an oscillating condensate with $n=8$ after
the trap energy is suddenly reduced to half at $T=0$.  The $|\psi(0)|^2$
for condensates corresponding to the initial and final traps are denoted
by the two straight lines.  The time evolution is performed using time
step $\Delta=0.03, \Delta_1 =0.01$, space step $h=0.0001$, and
$x_{\mbox{max}} = 15$.}

Next we consider two time-evolution problems using the time-dependent
method. We consider the ground state in the repulsive case with $cn=8$. In
the first problem, at $t=0$, the trap is suddenly removed.  In the second,
at $t=0$, the trap energy is suddenly reduced to half of the starting
value. In both cases we study how the system evolves with time by solving
the time-dependent GP equation using time step $\Delta = 0.03$,
$\Delta_1\sim 0.01$ and space
step $h=0.0001$.  Both these problems are intrinsic time-dependent
problems and can be studied numerically and experimentally.

The condensate cannot exist in the absence of the trap.  In the first case
after the trap is removed at $t=0$, the radius of the condensate increases
and the wave function extends over a larger region of space. We solve the
time-dependent GP equation at different times. In Fig. 3 we plot the wave
function at different reduced times $T = t/0.03$. The condensate
increases in size monotonically with time and eventually disappears. 

\vskip -3.8cm
\postscript{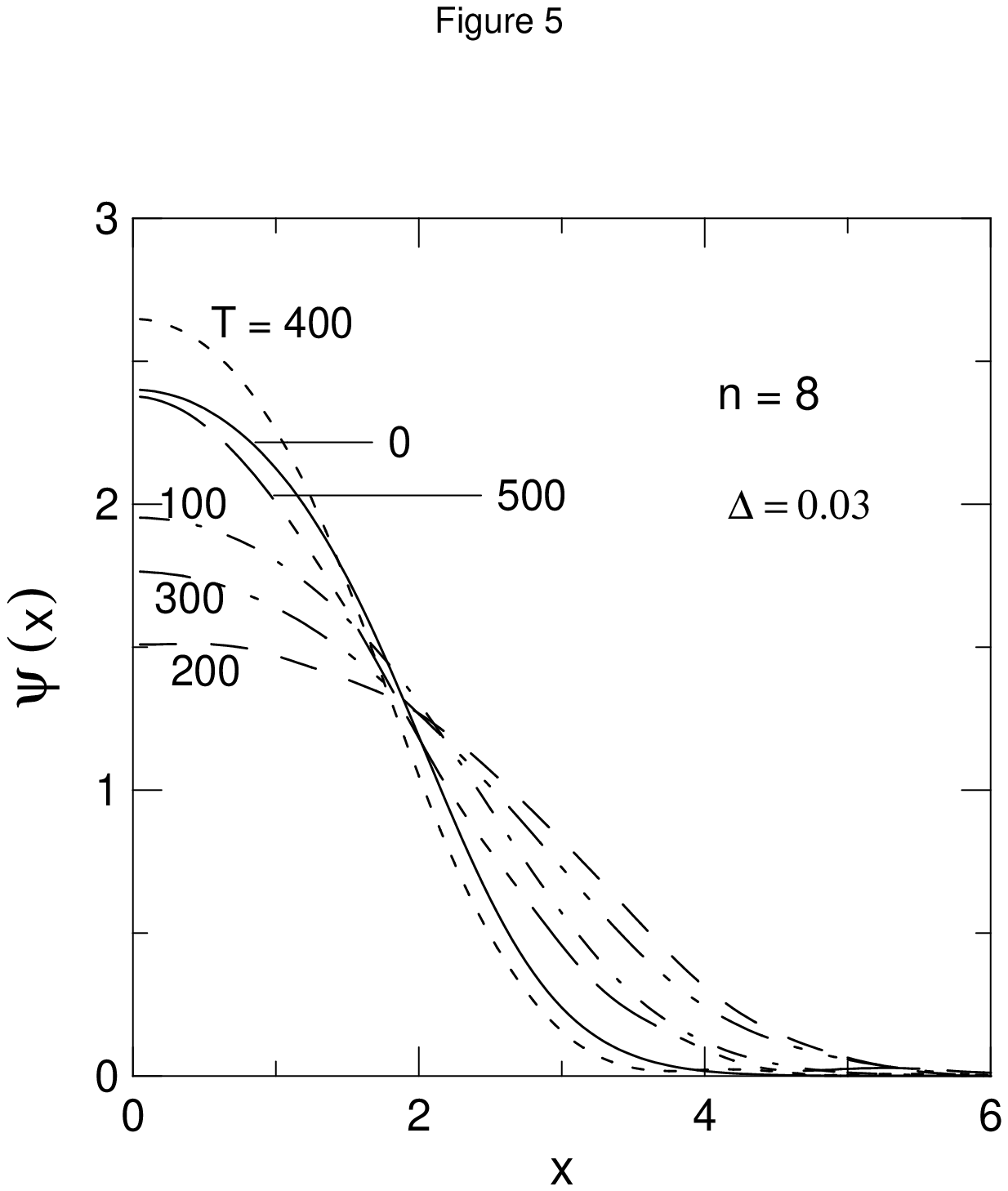}{1.0}    
\vskip -2.1cm

\vskip .4cm

{ {\bf Fig. 5.}
Condensate wave function $\psi(x)$ at different times $T=t/0.03$ for an
oscillating condensate after the trap energy is suddenly reduced to half
at $T=0$. All parameters are the same as in Fig. 3. }

In the second problem at $t=0$, we reduce the trap energy suddenly to half
of the initial value corresponding to a stable final configuration for the
condensate in the repulsive case. The system is now found to oscillate
between the initial and
final stationary states. In the absence of the nonlinearity, the system
executes sinusoidal oscillations between the two stable configurations.
However, in this nonlinear problem the system executes oscillations with
evergrowing amplitude. To illustrate this oscillation we plot in Fig. 4
the the central probability density $|\psi(0)|^2$ versus reduced time $T =
t/0.03$.  The $|\psi(0)|^2$ for condensates corresponding to the initial
and final trap energies are denoted by the two straight lines. We see that
the oscillation increases with time. In our numerical study we find that
after a very large number of iterations (several thousands) the amplitude
may become very large. However, we do not know if this result makes sense
physically as the cumulative numerical error of the type shown in Fig. 2
will also grow after a very large number of iterations, which will
possibly invalidate our conclusion. However, the solution presented in
Fig. 4 is stable numerically and is the acceptable physical solution of
the problem after a small number of iterations.  This interesting behavior
can possibly be observed experimentally and deserves further theoretical
and numerical studies. In Fig. 5 we plot the wave functions of the system
at different times which have very acceptable and smooth behavior. As the
number of particles of the system continues fixed, the wave functions of
smaller amplitudes have larger spacial extension [mean square radius
(\ref{7})] so that the normalization condition (\ref{5}) is preserved.

\section{Summary}

In this paper we present a numerical study of the Gross-Pitaevskii
equation for BEC in two space dimensions under the action of a harmonic
oscillator trap potential for bosonic atoms with attractive and repulsive
interparticle interactions using time-dependent and time-independent
approaches. Both approaches are used for the study of the stationary
problem. In addition some evolution problems were studied by the
time-dependent approach.  We derive the boundary conditions (\ref{10}) and
(\ref{11}) of the solution of the dimensionless GP equations (\ref{d}) and
(\ref{3}). These boundary conditions are used for the solution of the
stationary problem using both the time-dependent and time-independent
approaches. 

The time-dependent GP equation is  solved by discretizing it using a
Crank-Nicholson-type scheme, whereas the time-independent GP equation is
solved by numerical integration using the four-point Runge-Kutta rule. 
In both cases numerical difficulty appears for large nonlinearity
($cn>20$). For medium nonlinearity, the accuracy of the time-independent
method can be increased by reducing the space step $h$. However, the
solution of the time-dependent approach exhibits intrinsic oscillation 
with time iteration which is independent of  space and time steps used
in discretization. 

The ground-state wave function is found to be sharply peaked
near the origin for attractive interatomic interaction. For a repulsive
interatomic interaction the wave function extends over a larger region of
space.  In the case of an attractive potential, the mean square radius
decreases with an increase of the number of particles in the condensate. 
Consequently, a stable solution of the GP equation can be obtained for a
maximum number of particles in the condensate as given in Eq.  (\ref{12}).

In addition to the stationary problem we studied two evolution problems
using the time-dependent approach. A stable bound state is considered and
the trap potential is suddenly removed or reduced to half at $t=0$. If the
trap is removed suddenly, the system gradually and monotonically increases
in size with time and eventually it disappears occupying the whole space
with zero density.  If the trap energy is suddenly reduced to half, the
system oscillates around the two stationary positions. The amplitude of
the oscillation continue to increase with time. This behavior is
interesting and can be studied experimentally in the future.

The work is supported in part by the Conselho Nacional de Desenvolvimento
Cient\'\i fico e Tecnol\' and Funda\c c\~ao de Amparo \`a Pesquisa do
Estado de S\~ao Paulo of Brazil.

\end{document}